\documentclass{IEEEtran}

\usepackage{array}
\usepackage{amssymb}
\usepackage[cmex10]{amsmath}
\usepackage{nicefrac}
\usepackage{epsfig}
\usepackage{epstopdf} 

\usepackage{latexsym}
\usepackage{times}
\usepackage{url}
\usepackage{graphicx}

\begin{document}

\title{A Robust Low-Complexity MIMO Detector for Rank 4 LTE/LTE-A Systems }
\author{\IEEEauthorblockN{Shashi Kant$^{(\dagger)}$, Fredrik Rusek$^{(\dagger,\ddagger)}$, and Basuki E. Priyanto$^{(\dagger)}$}\\
\IEEEauthorblockA{$^{(\dagger)}$Huawei Technologies Sweden AB, 22369 Lund, Sweden.}\\
\IEEEauthorblockA{$^{(\ddagger)}$Dept. of Electrical and Information Technology, Lund University, 22100 Lund, Sweden.}\\
\{firstname.middlename.lastname\}@huawei.com
}

\maketitle

\newcommand{\sk}[1]{{\color{red} #1}} 
\newcommand{\bep}[1]{{\color{blue} #1}} 
\newcommand{\fr}[1]{{\color{green} #1}}
\newcommand{\comments}[1]{{\color{magenta} #1}}

\iftrue
\renewcommand{\vec}[1]{\ensuremath{\boldsymbol{#1}}}
\newcommand{\herm}{{\rm H}}
\newcommand{\expect}{\mathbb{E}}
\newcommand{\C}{\mathbb{C}}
\newcommand{\be}{\begin{equation}}
\newcommand{\ee}{\end{equation}}
\newcommand{\pdf}{\mathcal{P}}
\fi

\def\baselinestretch{1}

\begin{abstract}
This paper deals with  MIMO detection for rank 4 3GPP Long-Term-Evolution (LTE) systems.  The paper revolves around a previously known detector \cite{i_lee_mmse_rd_mls_april_2010}, which we shall refer to as RCSMLD (Reduced-Constellation-Size-Maximum-Likelihood-Detector). However, a direct application of the scheme in \cite{i_lee_mmse_rd_mls_april_2010} to LTE/LTE-A rank 4 test cases results in unsatisfactory performance. The first contribution of the paper is to introduce several modifications that can jointly be applied to the basic RCSMLD scheme which, taken together, result in excellent performance. Our second contribution is the development of a highly efficient hardware structure for RCSMLD that allows for an implementation with very few multiplications. 
\end{abstract}

\section{Introduction}
Multiple-input Multiple-output (MIMO) in spatial multiplexing (SM) mode is a key technology in emerging systems to meet high data rate requirements. LTE User Equipment (UE) category 5 with a maximum data rate of 300 Mbps requires the usage of 4 spatial layers.  The MIMO detection problem has spurred an impressive amount of research, and a good overview can be found in, e.g., \cite{Lspm}. However, many detectors that are considered "`low-complexity"' in academia are not easily implementable in LTE UE chipsets at the time being. An example is sphere decoding (SD), which suffers from a variable complexity. This requires the hardware design to take significant height so that the worst case can be accommodated - a design that leaves the chip idle most of the time. As a remedy, the fixed-complexity sphere decoder (FCSD) was proposed in \cite{FCSD}, but we have observed that the FCSD has in general a too weak trade-off between performance and complexity. Another example is lattice-aided reduction (LAR) techniques, which is based upon the LLL-algorithm. However, the complexity of the LLL is high and not constant. Many detectors are based upon the K-best principle for searching a tree. However, as the constellation is large, say, 64-QAM, the branching factor of the tree is large which yields high complexity even when a small number of nodes per depth is stored.

In this paper, we deal with  downlink MIMO detection for  rank 4 SM schemes. We have the following constraints in mind (i) the complexity is constant, (ii) hardware complexity is low and can efficiently be pipelined, (iii) performance is as good as full search, and (iv) maintain good performance over \textit{all} supported modulation and coding schemes in \textit{all} 3GPP-like test scenarios.
This paper summarizes our efforts to design an "`implementable"' detector that meets the above constraints. The work can easily be adopted to any rank $\leq$ 4; but extensions to rank $>$ 4 is possible with modifications--not treated here.  

\subsection{Notation} The $i$-th element of the vector $\vec{a}$ is denoted by ${a_i}$.  The element in the $i$-th row and $j$-th column of the matrix $\vec{A}$ is denoted by $\vec{A}\left[i,j\right]$, and its conjugate transpose by $\vec{A}^{\herm}$. The $m$-th column of a matrix $\vec{A}$ is denoted by $\vec{a}_m$. The complex conjugate to a scalar variable $b$ is denoted by $b^*$. The $K \times K$ identity matrix is written as $\vec{I}_K$. The expectation operator is denoted by $\expect\{ \cdot \}$; $\mathbb{C}$ is the set of complex numbers. The real part of a complex number $x$ is denoted by $\mathcal{R}\{x\}$.

\section{System Model} \label{sec:system_model}
We consider a single-cell scenario where the e-NodeB is equipped with $N_{\rm T}$ transmit (Tx) antennas, and the UE is equipped with $N_{\rm R}$ receive (Rx) antennas. The SM transmission scheme employs $N_{\rm L} \leq \min\left\{N_{\rm T},N_{\rm R}\right\}$ spatial layers such that the rank is  $N_{\rm L}$. 

Under the assumption that the UE is perfectly synchronized with the serving-cell, the received frequency-domain complex signal vector for a resource-element (RE) carrying physical downlink shared channel (PDSCH) within a subframe reads, 
\be \label{eqn:general_signal_model_per_re}
    \vec{y}\, = \vec{H} \vec{x}\, + \, \vec{w} \, , 
\ee
where $\vec{H}\in \C^{N_{\rm R} \times N_{\rm L}}$ describes the channel matrix; $\vec{x}= {{{\rm mapping}}}\left[b_1,b_2,\ldots,b_{QN_{\rm L}}\right] \in \mathcal{S}^{N_{\rm L}}$ denotes the transmitted data symbol vector which is a mapping of the coded-bit vector $\left[b_1,b_2,\ldots,b_{QN_{\rm L}}\right]$, where each element of the mapped data symbol $\vec{x}$ belongs to a finite-alphabet set $\mathcal{S}$ corresponding to a $2^Q$-QAM constellation. Although in LTE, the modulation alphabet can be different for different code-words,  we assume the same modulation alphabet for both code-words without loss of generality. The vector $\vec{w}$ is complex  Gaussian noise  with covariance matrix $N_0 \vec{I}_{N_{\rm R}}$.

The optimal (bit-wise) maximum a-posteriori (MAP) detection of the $i$-th transmitted code-bit is obtained by computing the log-likelihood ratios (LLRs) as
\begin{align}
L_{\rm MAP}^{i} &= \log \left(\frac{ \sum\limits_{\vec{x} :b_i(\vec{x})=1}  \exp \left( - \frac{\mu(\vec{x})}{N_0}  \right) \, }{ \sum\limits_{\vec{x} :b_i(\vec{x})=0}  \!\exp \!\left( -\frac{\mu(\vec{x})}{N_0}   \right)  } \right) \label{eqn:full_log_APP_with_prior}
\end{align}
where $\mu(\vec{x})=\|\vec{y}-\vec{Hx}\|^2$.

To reduce the complexity of the log-MAP approach, one resorts to the max-log-MAP (MLM) approach by replacing the summation in (\ref{eqn:full_log_APP_with_prior}) with the maximum term which yields
\be
 L_{\rm MLM}^{i} =  -\max \limits_{\vec{x} :b_i(\vec{x})=1} \left\{\mu(\vec{x})/N_0\,  \right\} + \max \limits_{\vec{x} :b_i(\vec{x})=0} \left\{ \mu(\vec{x})/N_0  \right\} \label{eqn:full_MLM_with_prior}  .
\ee

\subsection{The Basic RCSMLD Scheme} \label{sec:proposed_detectors}

In this section we lay down the operations of the basic RCSMLD in order to establish a framework for the proposed improvements.
The RCSMLD in \cite{i_lee_mmse_rd_mls_april_2010} first filters the received signal $\vec{y}$ with the MMSE filter to obtain
\be \label{mmse}\hat{\vec{x}}=\vec{H}^{\rm H}(\vec{H}\vec{H}^{\rm H}+N_{\rm 0}\vec{I})^{-1}\vec{y}.\ee
For each element $\hat{x}_k$ of $\hat{\vec{x}}$, we construct a set $\mathcal{C}_k$ of cardinality $M_k$ that contains the constellation points in $\mathcal{S}$ that lie the closest to $\hat{x}_k$. The cardinality of the set is pre-defined in order to have a fixed complexity. In other words, $\mathcal{C}_k$ is constructed such that
$$s\in \mathcal{C}_k,\; \tilde{s}\notin \mathcal{C}_k \Rightarrow |\hat{x}_k-s|^2<|\hat{x}_k-\tilde{s}|^2,\;{\rm and}\;|\mathcal{C}_k|=M_k.$$
The rationale behind the construction of these sets is that the MMSE estimate $\hat{\vec{x}}$ is a fairly good estimate of the transmitted symbol vector $\vec{x}$. Therefore, it is unlikely that a vector far away from $\hat{\vec{x}}$ is the solution to (\ref{eqn:full_MLM_with_prior}). Thus, the Cartesian product of the sets, i.e., $\mathcal{C}=\mathcal{C}_1\times \cdots \times \mathcal{C}_{N_{\rm L}}$ is a likely subset of the full search-space $\mathcal{S}^{N_{\rm L}}$ for the solution to (\ref{eqn:full_MLM_with_prior}) to lie in. The RCSMLD is then completed by replacing (\ref{eqn:full_MLM_with_prior}) with the simpler
\be\label{bml}
 L_{\rm RCSMLD}^{i} = \max \limits_{\vec{x} \in {\mathcal{C}:{{b_i(\vec{x})} = 1}}}\! \left\{\! - {\frac{{ {\mu(\vec{x})}}}{{N_0}} } \!\right\}-\max \limits_{\vec{x} \in {\mathcal{C}:{{b_i(\vec{x})} = 0}}}\! \left\{\! - {\frac{{ {\mu(\vec{x})}}}{{N_0}} } \!\right\}.																			
\ee
In some cases one of the sets to maximize over may be empty, which causes a "`missing bit"' problem. This we shall handle separately and we get back to this later in the paper. Altogether, the RCSMLD has a complexity that is given by $M=\prod_k M_k$ and this complexity can significantly be smaller than the complexity of the full search $|\mathcal{S}|^{N_{\rm L}}$ without any major performance degradation. Yet, when we apply the RCSMLD scheme to LTE rank 4 test cases, unacceptably large values of $M$ are needed in order for the RCSMLD to reach close to MLM-detection results.

\section{Proposed Modifications to RCSMLD}
In this section  we make several modifications that substantially improves the performance/complexity trade-off. Due to lack of space, we only mention some of the  modifications briefly, and defer full details to a future extended version of the paper; but more information can be found in \cite{patent1}.
\subsection{Improved Construction of the Candidate Sets $\mathcal{C}_k$ based on Interference Cancellation} \label{imo}
The main problem with the basic RCSMLD is that the construction of the sets $\mathcal{C}_k$ by the pure MMSE filtering (\ref{mmse}) is quite weak and should be replaced by a better detection technique having low-complexity in order to fulfill an overall complexity budget. We propose to use an iterative MMSE technique where a soft-parallel interference cancellation (MMSE-SPIC) step is performed within each iteration by utilizing only the demodulation soft-bits\footnote{One could also utilize the soft-bits from the Turbo decoder. But, we do not consider this in order to keep the complexity budget and latency low.}. Let $N_{\rm iter}$ denote the total number of iterations of the MMSE-SPIC. Then, the operations in each iteration are as follows (e.g., cf. \cite{Cstuder_mmse_spic})

\begin{enumerate}
		\item For all $n$, compute  $\breve{x}_n\!=\!\expect{\left\{x_n\right\}}$ and $\sigma_{x_n}^2\!=\!\expect \left\{\left|x_n\!-\!\breve{x}_n\right|^2\right\}$ using the LLR outputs $\left\{L_{\rm MMSE}^{i}\right\}$ from the MMSE demodulator as prior information. In the first iteration $L_{\rm MMSE}^{i}\!=\!0,$ so that $\breve{x}_n\!=\!0$ and $\sigma_{x_n}^2\!=\!1, \forall n$.

\item For each layer $n$,  perform PIC:
\be
  \widetilde{\vec{y}}_{\left(n\right)} = \vec{y} - \sum \limits_{m=1,m\neq n}^{N_{\rm L}} \vec{h}_m \breve{x}_m. 
\ee

		\item Apply the MMSE filter to each signal $\widetilde{\vec{y}}_{\left(n\right)}$ such that the $n$-th layer output reads
		$\hat{x}_n= \vec{g}_n^{\herm}\widetilde{\vec{y}}_{\left(n\right)}$
		where the MMSE filter $\vec{g}_n^{\herm}$ is the $n$-th row vector of 
$
  \vec{G}^{\herm} = \left(\vec{H}\vec{H}^{\herm}\vec{R}_{xx} + N_0 \vec{I}_{N_{\rm L}}\right)^{-1} \vec{H}^{\herm} \, ,
$
 and the diagonal matrix $\vec{R}_{xx} = {\rm diag}\left(\sigma_{x_1}^2,\ldots,\sigma_{x_n}^2,\ldots,\sigma_{x_{N_{\rm L}}}^2\right)$.
\item Invoke the Gaussian approximation to the estimate
\begin{align}
  \pdf\left(x_n\left|\widehat{x}_n\right.\right)
	                         \propto  \exp\left(\frac{-\left|\widehat{x}_n-\beta_n x_n\right|^2}{\sigma_{\widetilde{w}_n}^2}\right) \label{eqn:posterior_pdf_x_mmse_pic} , 
\end{align}
where, $\sigma_{\widetilde{w}_n}^2 = \beta_n \left(1 - \sigma_{x_n}^2{\beta_n} \right)$ is the post-processing noise-plus-interference variance,  $\beta_n = \vec{g}_n^{\herm} \vec{h}_n$. The post-processing signal to noise-plus-interference ratio (SINR) for $n$-th layer reads $\nicefrac{\beta_n^2}{\sigma_{\widetilde{w}_n}^2}$.

\item Compute the set of LLRs $\left\{L_{\rm MMSE}^{i}\right\}$ as\footnotemark[2]
\begin{align}
 L_{\rm MMSE}^{i} &= \max \limits_{x_n\in \mathcal{S}:b_i(x_n) = 1}\; \left(\frac{-\left|\widehat{x}_n-\beta_n x_n\right|^2}{\sigma_{\widetilde{w}_n}^2}\right) \notag \\ & - \max \limits_{x_n\in {\mathcal{S}:b_i(x_n)} = 0}\; \left(\frac{-\left|\widehat{x}_n-\beta_n x_n\right|^2}{\sigma_{\widetilde{w}_n}^2}\right) \label{eqn:uyt}.																			
\end{align}
\footnotetext[2]{We  abused the notation by writing $b_i(x_n)$; by this we mean that the $i$th bit of a vector $\vec{x}$ is already known to fall within element $n$ of the vector.
}

\end{enumerate}
Once $N_{\rm iter}$ iterations of the steps 1) -5) above has been carried out, the sets $\mathcal{C}_k$ can be constructed  by choosing its members as the $M_k$ constellation points in $\mathcal{S}$ that maximize (\ref{eqn:posterior_pdf_x_mmse_pic}).

A suitable choice is to use $N_{\rm iter}=2$ or $3$. The total number of candidates $M$ can significantly be lowered compared with using a pure MMSE step (i.e., $N_{\rm iter}~=~1$), and this reduction far exceeds the complexity increase of adopting an iterative construction of the candidate sets.

\subsection{Linear Combination}\label{lc}
The computed LLRs from the RCSMLD (\ref{bml}) are not optimal, no matter whether the improved candidate generation technique from Section \ref{imo} was used or not. 
A simple, yet highly effective, improvement is to apply a linear combining of the LLRs obtained from RCSMLD and MMSE-SPIC as
\begin{align}
	L_{\rm demod}^{i} = \alpha L_{\rm RCSMLD}^{i} + \left(1-\alpha\right)L_{\rm MMSE}^{i} \quad \forall i,
\end{align}
where $L_{\rm MMSE}^{i}$ is the obtained LLR from the last iteration of (\ref{eqn:uyt}). The value of $\alpha$ is obtained via simulations and $\alpha=0.5$ works well. Linear combination solves also the missing bit problem. If $L_{\rm RCSMLD}^{i}$ is undefined for some $i$, one can set $\alpha=0$ so the LLR from the MMSE-SPIC is taken as output. This is less harmful than it appears, as an undefined $L_{\rm RCSMLD}^{i}$ implies that one is already very certain about bit $i$.

\subsection{Candidate Reduction}\label{cr}
One can reduce the number of candidate vectors $M$ by deleting the most unlikely set of candidates. To be specific, we  remove vectors from the set $\mathcal{C}$ which involves two or more of the least-likely candidates per layer. 

For example, consider a rank 3 scenario and let the number of candidate vectors to be selected be $[M_1,M_2,M_3]=[3,2,2]$, i.e., we have $M=\left|\mathcal{C}\right|=12$ candidate vectors to evaluate the metric of. Then, we never evaluate the metric of a candidate vector that contains 2 or more of the least likely symbols per spatial layer. For the given example, we remove 5 out of 12 vectors. The rationale is that the subset $\mathcal{C}$ of candidate vectors to consider becomes more "`sphere like"' as we removed  "`corners"' of it.
The candidate reduction method works  particularly well in low spatial correlation.

\subsection{A Real-Valued Formulation} \label{realsumis}
So far, the sets $\mathcal{C}_k$ contained complex valued symbols from the set $\mathcal{S}$. The main idea in this modification is to let the members of $\mathcal{C}_k$ be pairs where each pair contain one real-or-imaginary part from some layer $k$ and one real-or-imaginary part from some other layer $k^\prime$. This is accomplished by first expressing the system with with an entirely real-valued model, followed by a permutation of the columns of the real-valued channel matrix. In the initial MMSE step described in Section \ref{imo}, all processing is performed jointly on pairs of real symbols. This strategy results in more degrees of freedom in the design of the detector, and leads to impressive gains. Due to lack of space, we defer further details to an extended version of this paper, but numerical results will be presented and more information can be found in \cite{patent1}.

\subsection{Channel Estimation Error Aware RCSMLD}\label{ceaware}
In practice the channel matrix $\vec{H}$ is never perfectly known. A more realistic system model than (\ref{eqn:general_signal_model_per_re}) is
\be 
\vec {y} = {\vec{H}}\vec{x}+\vec{Ex} + \vec {w},
\ee 
where $\vec{E}$ is the channel estimation error matrix and $\vec{H}$ is  the estimated channel matrix by the UE. We assume that all elements in $\vec{E}$ are IID, and that they follow a zero-mean complex Gaussian distribution with variance $\sigma_{\rm ce}^2$.
Moreover, we assume that $\sigma_{\rm ce}^2$ is known to the UE. With that, the noise density depends on the magnitude of $\vec{x}$, so that when evaluating the metric of $\vec{x}$, one should replace $\mu(\vec{x})/N_0$ by
$$N_{\rm R}\cdot\log\left({N_0+\|\vec{x}\|^2\sigma_{\rm ce}^2} \right) + \frac{\mu(\vec{x})}{N_0+\|\vec{x}\|^2\sigma_{\rm ce}^2}.$$

This change, that can lead to substantial performance increase \cite{Goldsmith}, can seamlessly be incorporated into RCSMLD directly in (\ref{bml}). For many other detectors, such as tree-searches, this is not the case as the total vector $\vec{x}$ is not available at intermediate depths in the search tree so that the total noise density $N_0+\|\vec{x}\|^2\sigma_{\rm ce}^2$ is not known. In RCSMLD, on the other hand, all candidate vectors are known \textit{prior} to computing any metrics, so that $N_0+\|\vec{x}\|^2\sigma_{\rm ce}^2$ is at hand. In the initial candidate generation step in Section \ref{imo}, $\|\vec{x}\|^2$ is not available, but a suitable remedy is to replace $N_0$ with $N_0+\mathbb{E}\{\|\vec{x}\|^2\}\sigma_{\rm ce}^2$.

\subsection{Markov Chain Monte Carlo (MCMC) Methods} \label{mcmc}
The RCSMLD scheme can nicely be merged with MCMC \cite{r_chen_b_mcmc_feb_2010}. We propose to introduce the Gibbs sampler after the initial  candidate generation in Section \ref{imo}. The motivation for considering MCMC is to reduce the number of candidates even further and to generate a better candidate list before computing their metrics in (\ref{bml}). Unlike methods proposed in the literature which utilize costly techniques to initialize the Gibbs sampler, e.g., QRD-M \cite{r_peng_qrdm_mcmc_2008}, fixed-sphere-decoders \cite{f_li_yuan_fsd_mcmc_2011}, our method utilizes  the low-complexity MMSE-SPIC technique in Section \ref{imo} to initialize multiple Gibbs samplers with multiple candidate vectors. 

As per our numerical results, the number of Gibbs sampler iterations per hypothesis candidate vector that can render good performance is typically two or three. Further details about the complexity-performance tradeoff by introducing MCMC is defered to a future extended version of the paper. 
\section{Efficient Hardware Structure to compute (\ref{bml})}
We next  turn our attention to an efficient way to compute all $M$ metrics $\mu({\bf x})=\|\vec{y}-\vec{Hx}\|^2$.
The value $\mu({\bf x})$ can be written as
\begin{align} 
    \mu({\vec{x}}) &\propto-\mathcal{R}\left\{{\vec{ x}}^{\herm} {\vec{H}}^{\herm} {\vec{ y}}\right\}+\frac{1}{2}{\vec{ x}}^{\herm}{\vec{ H}}^{\herm}{\vec{ Hx}} \notag \\
		                     &=\sum_{n=1}^{N_{\rm L}} |x_n|^2\frac{\vec{G}[n,n]}{2}-\mathcal{R}\left\{x_n^\ast \!\left[z_n\!\!-\!\!\sum_{m=1}^{n-1} \vec{G}[n,m]x_m\right]\right\}
\end{align}
where we have defined ${\vec G}={\vec H}^{\herm} {\vec H}$ and ${\vec z}={\vec H}^{\herm}{\vec y}$. 
The metric $\mu({\bf x})$ can now be evaluated in a recursive fashion over a tree structure. In the above summation, the index $n$ has the meaning of "tree-depth". This is so since for each stage $n$, only the symbols $x_n,x_{n-1},\ldots,x_1$ are involved in the computations. To better visualize this, it is useful to define the partial metric 
$$\mu_L(\vec{x})\triangleq \sum_{n=1}^L |x_n|^2\frac{\vec{G}[n,n]}{2}-\mathcal{R}\left\{x_n^\ast \!\left[z_n\!\!-\!\!\sum_{m=1}^{n-1} \vec{G}[n,m]x_m\right]\right\}$$ 
where $\mu(\vec{x})=\mu_{N_{\rm L}}(\vec{x})$  is the quantity of interest.
Then, we can directly reach the recursive formulation
\begin{align}\mu_L(\vec{x})&=\mu_{L-1}(\vec{x}) +|x_L|^2\frac{\vec{G}[L,L]}{2}\notag \\
&\quad -\mathcal{R}\left\{x_L^\ast \!\left[z_L\!-\!\sum_{m=1}^{L-1} \vec{G}[L,m]x_m\right]\right\}.
\end{align}
By making the additional definitions
\begin{align}
\gamma_k(x)&\triangleq-\mathcal{R}\{x^\ast z_k\}+|x|^2\frac{\vec{G}[k,k]}{2}\notag \\
\delta_{mn}(x,y)&\triangleq \mathcal{R}\left\{x^\ast \vec{G}[m,n]y\right\}.
\end{align}
we can express the recursive metric as
\be \label{rec} \mu_L(\vec{x})=\mu_{L-1}(\vec{x})+\gamma_L(x_L)+\sum_{m=1}^{L-1} \delta_{Lm}(x_L,x_m).\ee

The important observation is that the recursive formulation of the metric in (\ref{rec}) does not involve any multiplications at all whenever $\gamma_k(\cdot)$ and $\delta_{mn}(\cdot,\cdot)$ are computed. However, these functions needs only to be evaluated for the candidates in the sets $\mathcal{C}_k$.

To keep complexity low, the order in which one processes the layers is important. One should  order the spatial layers in ascending order based on the post-processing SINR of MMSE-SPIC such that the largest number of candidates is assigned to the weakest symbol layer (in the post-processing SINR sense) while the strongest symbol layer should be assigned the smallest number of candidates. For example, with 64-QAM inputs, the number of candidates per  layer can be represented in vector form $[M_1,M_2,M_3,M_4]=[14,9,5,4]$, which implies that the total number of candidates are $M=\left|\mathcal{C}\right| = 14 \times 9 \times 5 \times 4 = 2520$ (which corresponds to only $\approx\!0.015\%$ of a full search). The weakest layer has 14 candidate symbols, while the strongest layer considers only 4.

After straight-forward derivations, one can conclude that only
$$\underbrace{3\sum_{k=1}^{N_{\rm L}} M_k}_{\gamma\;\rm{part}}+\underbrace{2\sum_{k=2}^{N_{\rm L}} M_k\sum_{\ell=1}^{k-1}M_{\ell}}_{\delta\; \rm{part}}$$
real multiplications are needed. However, if the channel matrix $\vec{H}$ remains constant for some time, then the part related to $\delta$ only needs to be computed once per coherence interval.
Finally, we mention that the number of real additions to compute all the $M$ metrics is
$$2\sum_{k=1}^{N_{\rm L}} M_k+\sum_{k=2}^{N_{\rm L}} M_k\sum_{\ell=1}^{k-1}M_{\ell} + \sum_{\ell=3}^{N_{\rm L}} M_{\ell}\sum_{k=1}^{\ell-2}\prod_{n=1}^k M_k+2\sum_{k=2}^{N_{\rm L}} \prod_{\ell=1}^k M_{\ell}.$$

Let us consider a numerical example. Suppose that we have $[M_1,M_2,M_3,M_4]=[8,7,4,4]$. This gives a total of 453 real multiplications in order to compute the metrics of 896 complex vectors. The number of real additions is 2474. 

\section{Simulation Results} \label{sec:simulation_results}
We next turn to an elaborate simulation study of the offered performance of our improved RCSMLD scheme for a number of 3GPP-like test cases.
The simulation parameters and the investigated test scenarios are summarized in Table \ref{tasim}.

\begin{table}
\begin{center}
\caption{Simulation Parameters for LTE PDSCH.}\label{tasim}\vspace{4mm}
\vspace*{-4mm}
\scalebox{0.7}{
	\begin{tabular}{|l||c|c|c|c|} \hline
	\textbf{Parameters}         		& \textbf{Test-1}   & \textbf{Test-2}                   & \textbf{Test-3} & \textbf{Test-4}  \\ \hline
	\multicolumn{1}{|l||}{BW (full-PRB allocation)}      & \multicolumn{4}{c|}{1.4 MHz}             \\ \hline
	\multicolumn{1}{|l||}{Tx EVM}         		 & \multicolumn{4}{c|}{6\%}                                                                     \\ \hline
	\multicolumn{1}{|l||}{MIMO Configuration}  & \multicolumn{4}{c|}{4$\times$4, Open Loop Spatial Mux. (TM3), fixed Rank=4.} \\ \hline
	\multicolumn{1}{|l||}{HARQ}                & \multicolumn{4}{c|}{8 processes and max. 4 transmissions.}                                   \\ \hline
	Modulation    		& 16QAM        & 64QAM                         & QPSK          & 16QAM       \\ \hline
	Code-rate     		& 0.72        &  0.85                        & 0.33         &  0.72      \\ \hline
	Channel Model               		& ETU300         & ETU70                          & ETU70        & ETU70     \\ \hline
	Correlation model               		&  Low        &  Low                         &  High       & UD11{\footnotemark[2]}    \\ \hline
	\multicolumn{1}{|l||}{Channel \& Noise-variance Estimate}     & \multicolumn{4}{c|}{Perfect}                                 \\ \hline
    \multicolumn{1}{|l||}{Other Information}         		 & \multicolumn{4}{c|}{LTE Turbo Encoder, control channel-4 OFDM symb.}   \\ \hline	
	\end{tabular}
}
\end{center}
\end{table}
\footnotetext[2]{UD means user-defined, i.e., non-3GPP, spatial correlation, whereby $\alpha=0.1$ and $\beta=0.1$ (cf. \cite{36101} for the definition of $\alpha$ and $\beta$).}

\begin{figure}[!htp]
\begin{center}
\scalebox{0.43}{\includegraphics{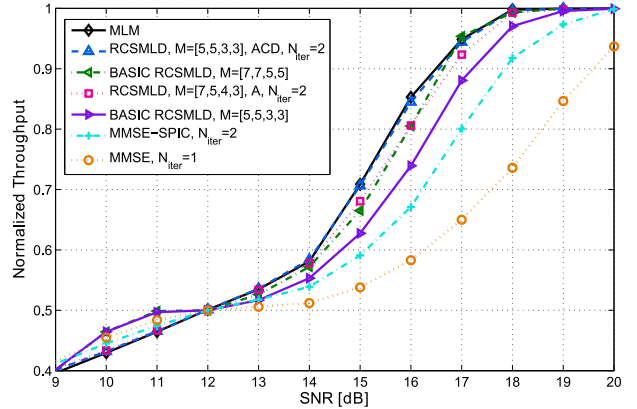}}
\end{center}
\vspace*{-4mm}
\caption{Simulation of Test case 1 with different detectors. Letters (e.g., XY) following "`RCSMLD"' refers to the proposed modifications in Sections III-X and III-Y, respectively.}
\vspace*{-4mm}
\label{f1}
\end{figure}

Results for Test 1 are given in Figure \ref{f1}. In this test we can see that the basic RCSMLD scheme from \cite{i_lee_mmse_rd_mls_april_2010} performs rather well and outperforms the MMSE detector with several dBs. However, there is still a gap to MLM, and to close this gap much larger $M_k$ values are needed. With the improved RCSMLD detector, the gap to MLM can be closed fully by using the improved candidate set construction ("`A"') in Section III-A and the real-valued formulation ("`D"') in Section III-D. Note that the improvement "`C"' is a complexity reduction improvement and not a performance improvement. We can also conclude that MMSE-SPIC alone is not competitive (increasing $N_{\rm iter}$ does not improve at all). The basic RCSMLD scheme with the $M$-vector $[7,7,5,5]$ evaluates 1225 metrics, while the improved RCSMLD with $[5,5,3,3]$ only evaluates 160 (after candidate reduction in Section III-C).

\begin{figure}[!htp]
\begin{center}
\scalebox{0.43}{\includegraphics{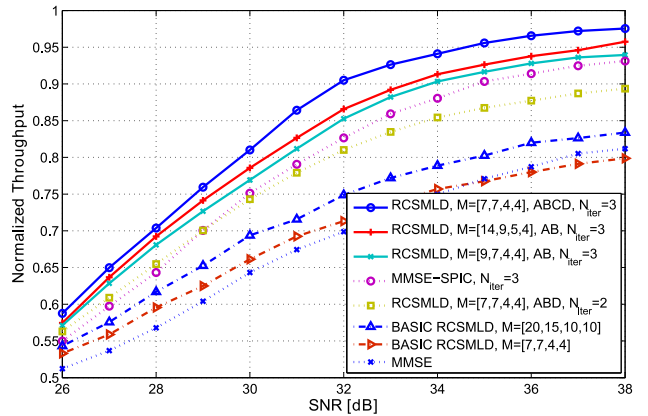}}
\end{center}
\vspace*{-4mm}
\caption{Simulation of Test case 2 with different detectors.}
\vspace*{-2mm}
\label{f2}
\end{figure}

Results for the more challenging Test 2 with 64-QAM inputs are given in Figure \ref{f2}. In this case we can see that the basic RCSMLD detector is far too weak, as it does not improve much over MMSE even with very large values of $M$. Since we have 64-QAM inputs, we have not been able to generate the MLM curve due to its high complexity. The gains of using the improvements can be seen clearly. Using the improvements from Sections III-A, III-B, III-C, and III-D, we can reach gains of several dBs over an MMSE-SPIC detector. The effectiveness of the improvments are clear as the smallest M-vector $[7,7,4,4]$, performs the best. This corresponds to computing the metrics of just 648 candidate vectors out of more than 16 millions.

\begin{figure}[!htp]
\begin{center}
\scalebox{0.43}{\includegraphics{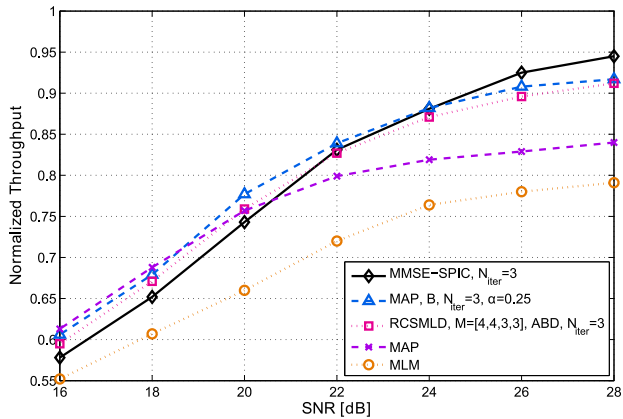}}
\end{center}
\vspace*{-4mm}
\caption{Simulation of Test case 3 with different detectors. }
\vspace*{-4mm}
\label{f3}
\end{figure}

In Figure \ref{f3} we show results for test case 3 which is based on QPSK inputs. Although the MLM detector is of low-complexity, so that complexity reduction is perhaps not critical, we report this case as interesting observations can be made. The two bottom curves at high SNR are for MAP and the MLM, i.e., the detectors (\ref{eqn:full_log_APP_with_prior}) and (\ref{eqn:full_MLM_with_prior}), which are usually thought of as optimal. However, large gains are possible by incorporating linear combination between the MMSE-SPIC output and the MAP output, as is shown with the dashed triangle-marked curve. In fact, at high SNR the MMSE-SPIC detector is the best among all the detectors.
Compared with the MAP approach, the improved RCSMLD, where the most important improvement is  the linear combination from Section III-B, has slightly reduced complexity both in terms of number of candidates and also since it uses the MLM computations rather than the MAP computation when generating LLRs. As can be seen, the RCSMLD is robust, but the linear combination factor $\alpha$ should be better tuned in order to not sacrifice performance at high SNR. At low SNR, one could use $0.5 < \alpha < 1$ and at high SNR $\alpha=0$ is more suitable for this test condition. Although at first glance, the fact that the MAP performs so badly may appear strange, our investigations show that the effect can partly be explained by the composite effect of the  high spatial correlation and transmit EVM.

\begin{figure}[!htp]
\begin{center}
\scalebox{0.43}{\includegraphics{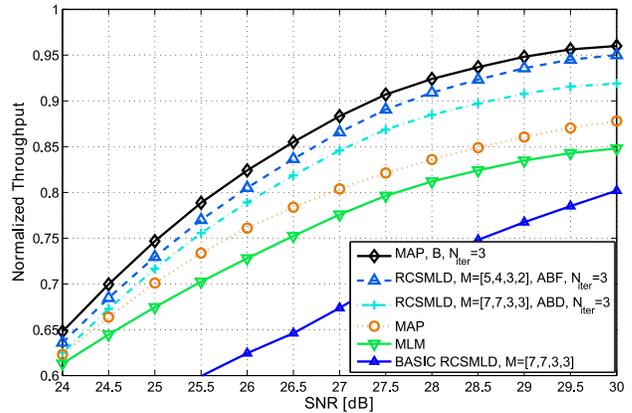}}
\end{center}
\vspace*{-4mm}
\caption{Simulation of Test case 4 with different detectors. }
\vspace*{-4mm}
\label{f4}
\end{figure}

Although not much details were given about the MCMC improvement, we report results for it in Figure \ref{f4} under the test case 4. Again, we can see that the basic RCSMLD scheme performs poorly, and that MAP and MLM without any linear combination with MMSE-SPIC are far from optimal. However, a linear combination of MAP with MMSE-SPIC results in excellent performance and is shown in the top curve. For this test case we made two tests with the improved RCSMLD, one with MCMC (3 Gibbs sampling iterations) and one without; in both cases it is critical to do linear combination with MMSE-SPIC. It can be seen that the MCMC improvement is highly effective and allows for a further reduction of the candidate numbers. With MCMC activated, we can reach close to the performance of MAP with linear combination.

\section{Conclusion} \label{sec:conclusion_future_work}
We have proposed a number of improvements to a MIMO detector studied in \cite{i_lee_mmse_rd_mls_april_2010}, including an efficient hardware structure for it. With the proposed modifications, the complexity-performance tradeoff is significantly improved. We have presented simulation results for a number of LTE/LTE-A rank 4 test cases.
We conclude that with the improvements made, very few metrics need to be calculated to obtain excellent performance. It is also critical to perform a linear combination between the output of the detector with the output of the simpler MMSE-SPIC detector. This holds even for the MAP detector.

\end{document}